\documentclass[useAMS,usenatbib]{mn2e}
\usepackage[final]{graphics}
\title[RR Pictoris: an old nova showing superhumps and QPOs]
      { RR Pictoris: an old nova showing superhumps and QPOs}
\author[L. Schmidtobreick, C. Papadaki, C. Tappert, A. Ederoclite]
       {L. Schmidtobreick$^{(1)}$\thanks{lschmidt@eso.org}, 
        C. Papadaki$^{(2,1)}$\thanks{cpapadak@vub.ac.be},
        C. Tappert$^{(3)}$\thanks{ctappert@astro.puc.cl}, 
        and A. Ederoclite$^{(1)}$\thanks{aederocl@eso.org}\\
        $^{1}$European Southern Observatory, Casilla 19001, Santiago 19, Chile\\
        $^{2}$Vrije Universteit Brussel, PLeinlaan 2, 1050 Brussels, Belgium \\
        $^{3}$Departamento de Astronom\'{\i}a y Astrof\'{\i}sica,
           Pontificia Universidad Cat\'olica, Casilla 306, Santiago 22, Chile}
\begin{document}
\date{Accepted xxxx. Received xxxx; in original form xxxx}
\pagerange{\pageref{firstpage}--\pageref{lastpage}} \pubyear{2005}

\maketitle

\label{firstpage}

\begin{abstract}
We present time--resolved V--photometry of the old nova RR\,Pic. 
Apart from the hump--like variability, the light curves show the 
strong flickering and 
random variation typical for RR\,Pic. We do not find any convincing 
evidence for the previously reported eclipse. The extrapolated eclipse phase coincides 
with a broad minimum, but comparing the overall shape of the light curve 
suggests that the eclipse should actually be located around phase 0.2.
The orbital period which we derive from these data agrees well with the old one, any uncertainty
is too small to account for the possible phase shift.
Apart from the 3.48\,h period, which is usually interpreted as the orbital one,
we find an additional period at $P=3.78$\,h, which we interpret as 
the superhump period of the system; 
the corresponding precession period at 1.79\,d is also present in the data.
We also find indications for the presence of a 13\,min 
quasi--periodic oscillation. 

\end{abstract}


\section{Introduction}
Classical novae are a subclass of 
cataclysmic variables (CVs), close interacting binary stars with a 
white dwarf primary receiving matter from a Roche--lobe--filling 
late--type star. They are distinguished by the observation of a
thermonuclear runaway outburst, the nova explosion.
As such, RR Pic was discovered by \citet{spen31} at maximum light
in 1925 and, although it was a slow nova, it is supposed to be in its
quiescence state by now. The orbital period of 0.145025 days \citep{vogt75}
places it just above the period gap and into the regime of the
SW\,Sex type stars. Indeed, RR\,Pic has been found to show several
observational features typical for SW\,Sex stars \citep{schm+03} and
can thus be regarded as a nova--like CV with very high mass transfer rate.

Vogt found the lightcurve dominated by a very broad hump, often
interrupted by superimposed minima. He explained this behaviour
by an extended hot spot region with an inhomogeneous structure.
\citet{haef+82} however, explained their own time--resolved photometric and
polarimetric observations together with radial velocity variations of the
He\,II (4686\,\AA) emission line \citep{wyck+77} by
suggesting the presence of an additional source of radiation in the disc
opposite of the hot spot. The presence of such an emission source was confirmed
via Doppler tomography by \citet{schm+03} and \citet{ribe+06}.
From high speed photometry, \citet{warn86} concluded that during
the 1970s (about 50\,yr after outburst) structural changes have taken
place in the system, resulting in a more isotropic  distribution of the
emitted radiation. In addition, he has found evidence for a shallow, irregular
eclipse, showing RR\,Pic to be a high inclination system. Note that no
signature of an eclipse had been found in the previous lightcurves.
In addition to the orbital period, \citet{kubi84} found a periodic modulation
in the optical with a 15\,min period. He interpreted this as the rotation of
the white dwarf and concluded that RR\,Pic is an intermediate polar.
\citet{haef+85} however, repeated the high time--resolved photometry on a
longer time--scale and could not find any sign of this short period.
Since no 15\,min period variation is found in X--ray measurements
\citep{beck+81} either,
they concluded that Kubiak's variation was more likely a transient event
in the disc rather than a rotating white dwarf. Also Warner's high-speed
photometry does not reveal any period other than the orbital one.

\citet{schm+05} compared radial velocity curves of different epochs 
and noticed a shift of about 0.1 phases of the radial velocity curves 
of data taken two years apart. They argue that this might be due to 
unstable emission sources in the accretion disc or might indicate a change
in the orbital period. To test these alternatives, we performed new
time--resolved photometry of RR\,Pic with the aim to determine the 
orbital period and look for a possible change that could account for the 
observed phase shift. These data and the results are presented in this paper.

\section{Data}
\begin{table}
\centering
\caption{\label{obstab} Summary of the observational details: Date \& UT 
at the start of the first exposure, the number of exposures, the 
individual exposure time, the covered orbital cycles and
the acquisition ID are given.}
 \begin{tabular}{@{}cccccc@{}}
 \hline
 Date & UT  & $\#_{\rm exp}$ & $t_{\rm exp}$ [s] & cycles & ID\\
 \hline
 2005-02-08 & 00:45:44 & 483 & 20 & 2.03 & 1\\
 2005-02-09 & 01:17:57 & 533 & 20 & 1.76 & 1\\
 2005-02-10 & 01:17:23 & 495 & 20 & 1.63 & 1\\
 2005-02-11 & 01:13:37 & 530 & 20 & 1.74 & 1\\
 2005-02-12 & 01:09:17 & 540 & 20 & 1.77 & 1\\
 2005-02-13 & 01:07:03 & 499 & 20 & 1.79 & 1\\
 2005-02-14 & 01:06:42 & 534 & 20 & 1.76 & 1\\
 2005-03-18 & 01:36:23 & 240 & 20 & 0.80 & 1\\
 2005-03-20 & 01:03:17 & 320 & 20 & 1.08 & 1\\
 2005-03-25 & 02:31:58 & 80  & 20 & 0.26 & 2\\
 2005-03-26 & 00:24:42 & 320 & 20 & 1.06 & 2\\
 2005-03-26 & 23:58:29 & 329 & 20 & 1.15 & 2\\
 2005-03-31 & 00:38:39 & 241 & 20 & 0.92 & 2\\
 2005-03-31 & 03:52:05 & 56  & 20 & 0.18 & 1\\
 2005-04-09 & 23:29:27 & 318 & 20 & 1.06 & 2\\
 \hline
\end{tabular}
\end{table}
The time resolved photometry was done using a V--filter in front of 
a 512x512 CCD mounted on the 1.0\,m SMARTS telescope at CTIO, Chile.
The data were
taken in 2005 between Feb 07 and April 10 and cover about 18 orbital cycles
with a time resolution of 40\,s. The details of the observations are given
in Table \ref{obstab}. The reduction was done with IRAF and included the 
usual steps of bias subtraction and division by skyflats.

Aperture photometry for all stars on the CCD field was computed using the
stand--alone version of DAOPHOT and DAOMASTER \citep{stet92}.
Differential light curves were established with
respect to an average light curve of those comparison stars, 
which were present on all frames and checked to be non--variable. 
While the original idea was to use the same comparison stars for all
epochs, we had to settle for two sets of comparison stars, as some of the
later data were taken with a different acquisition. 
The first set included five, the second set six comparison stars;
the two sets are distinguished by the acquisition IDs 1 and 2 in 
Table \ref{obstab}, corresponding finding charts are given in the appendix.  
The difference in the target's magnitudes between the two sets 
were established from three common stars as 0.31. The magnitudes of the 
second set were shifted accordingly.

\section{Results}
\subsection{The orbital period}
\begin{figure*}
\centerline{\resizebox{15cm}{!}{\includegraphics{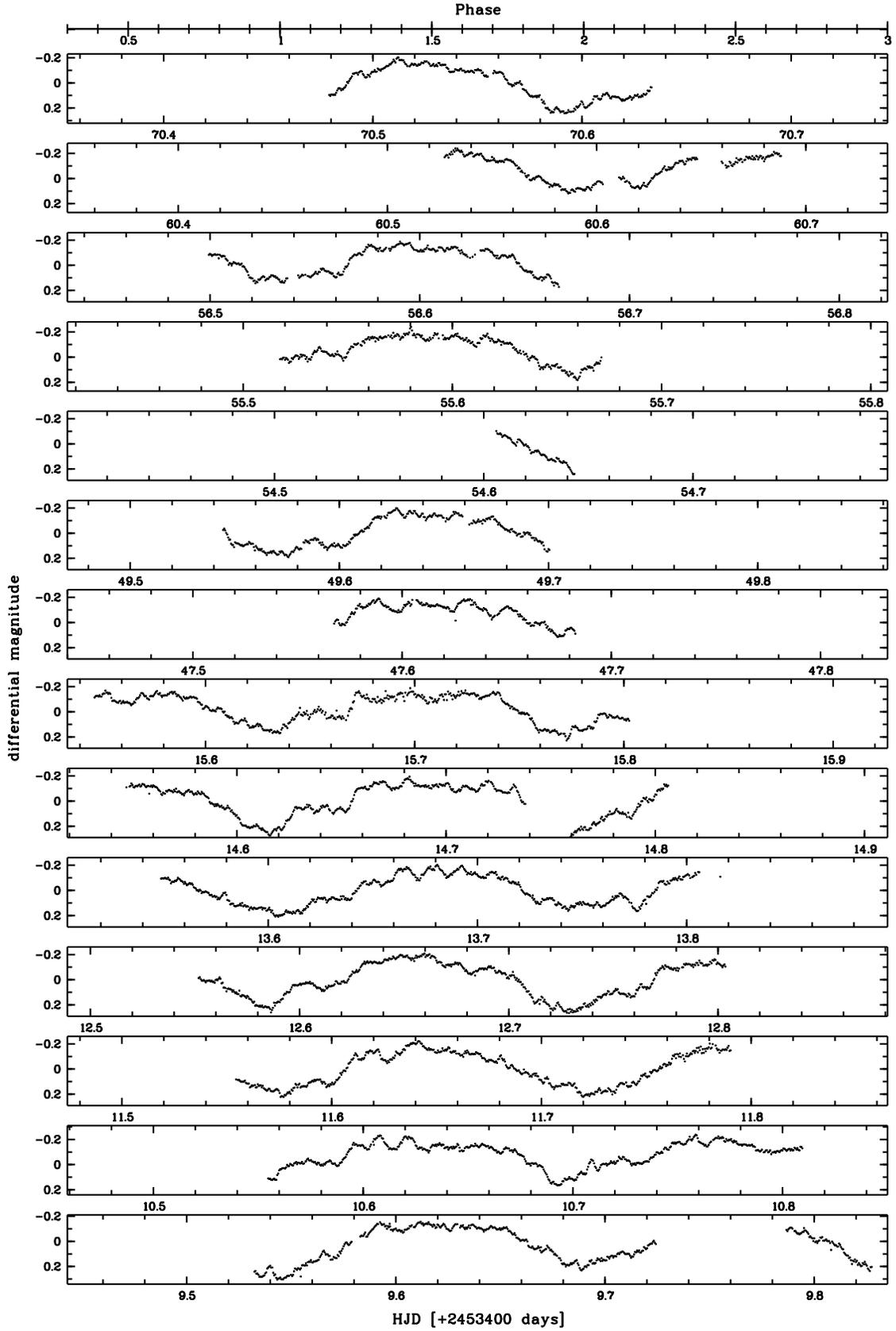}}}
\caption{\label{lc_all} The light curves of RR\,Pic. The phase refers to the
orbital period of $P=0.14502545(7)$\,d, $\phi =0$ corresponds to the 
eclipse--ephemeris as defined by \citet{schm+05}.}
\end{figure*}

\begin{figure}
\centerline{\rotatebox{-90}{\resizebox{!}{8.6cm}{\includegraphics{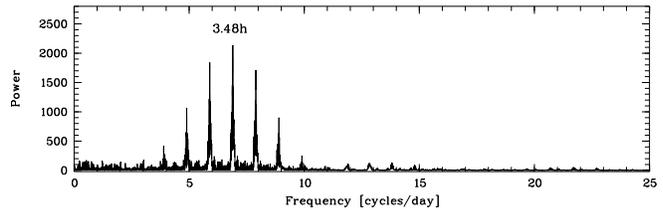}}}}
\caption{\label{scargle_lc_all} The Scargle--diagram for the data.
The period of 3.48\,h is clearly indicated by the maximum peak.}
\end{figure}

The obtained light curves are plotted in Figure \ref{lc_all}. They show a clear 
variation. The data were analysed with the Scargle and
analysis-of-variance (AOV) algorithms implemented in MIDAS 
\citep{scar82,schwa89} as well as with PERIOD\,04 \citep{lenz+05}. 
All three methods agree on the same period $P_{\rm ph} = 0.14503(7)$
which corresponds to a strong and unambiguous peak in the 
periodograms at $f_0 = 6.895\,\rm cycles/day$ 
(see Fig.\,\ref{scargle_lc_all}). 
We subdivided the observations in different sets but always obtained the same 
result, which thus indicates the robustness of this peak. 
We included photometric data taken in 2004 \citep{schm+05} to 
increase the accuracy of the orbital period and derived 0.1450255(1)d.
This value agrees very well with the formerly reported one
of $P=0.14502545(7)$\,d \citep{vogt75,kubi84}. 


We averaged our data with respect to the orbital period $P=0.14502545$\,d. 
The orbital phase was
computed for each data point using the ephemeris for the eclipse of
\citet{schm+05}. 
The data points have been averaged into
bins of 0.01 phases. On average, 55 individual data points went into 
each point of the average light curve. The result is plotted in 
Fig.\,\ref{ave_res}. For clarity, two orbital cycles are plotted (phase 0--2). 
For the first cycle, the sigma of each average point is indicated with an
error bar.
Apart from the broad hump, the light curve shows two clear features, 
a narrow peak on top of the hump at a phase of 0.42 and a sharp
minimum at a phase of 0.18. 
As expected for an average light curve, the flickering, which is present in
the individual light curves is completely gone.

With PERIOD\,04, we checked for the harmonics of the orbital period
and found peaks close to the theoretical values and down to the 6th 
harmonic at 0.0207\,d which has a power around 2 and is thus at the 
edge of detection. We used the theoretical harmonics
to fit the shape of the light curve. We then built an average fit 
in the same way as the average light curve.
The result is over--plotted in Fig.\,\ref{ave_res}, the residuals 
with the average light curve plotted below. Note that this best harmonic fit
also contains the above mentioned features: the narrow peak and the sharp
minimum. The residuals of fit and average light curve show high frequency
periodic variations with an amplitude around 0.01\,mag and harmonic to the
orbital one (as they would otherwise not appear in a phase diagram). 
This indicates, that harmonics of an order higher than 6 are present 
in the data even though they do not appear in the power spectrum.

We point out that neither the individual light curves of RR\,Pic nor the
average one show the eclipse that has been observed before. At phase 0, which
is the phase of the eclipse by \citet{warn86}, we find a broad minimum.


\subsection{The search for superhumps}

\begin{figure*}
\centerline{\resizebox{15cm}{!}{\includegraphics{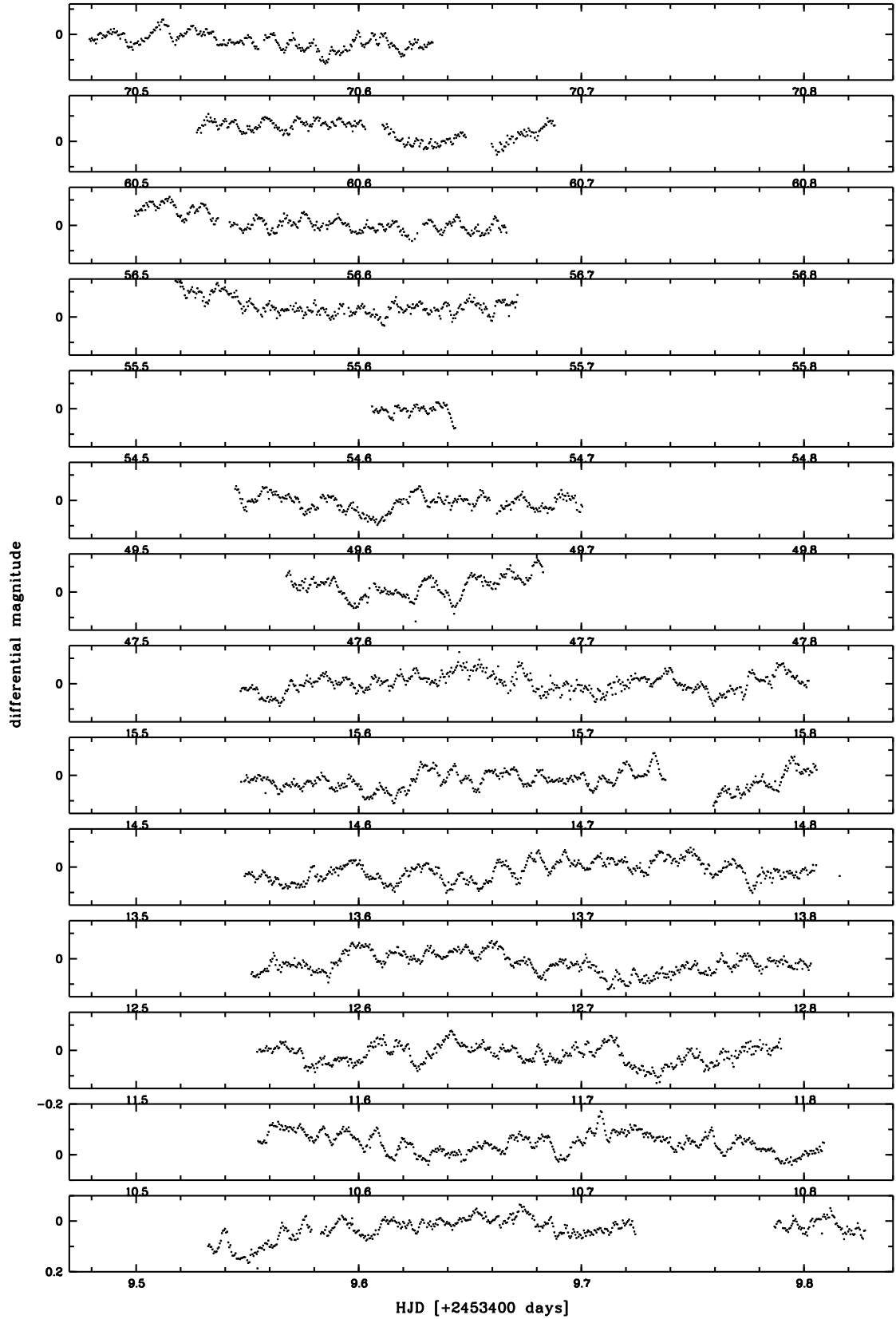}}}
\caption{\label{diff_all} The residual light curves of RR\,Pic, with the average
light curve subtracted. }
\end{figure*}

\begin{figure}
\rotatebox{-90}{\resizebox{!}{8.3cm}{\includegraphics{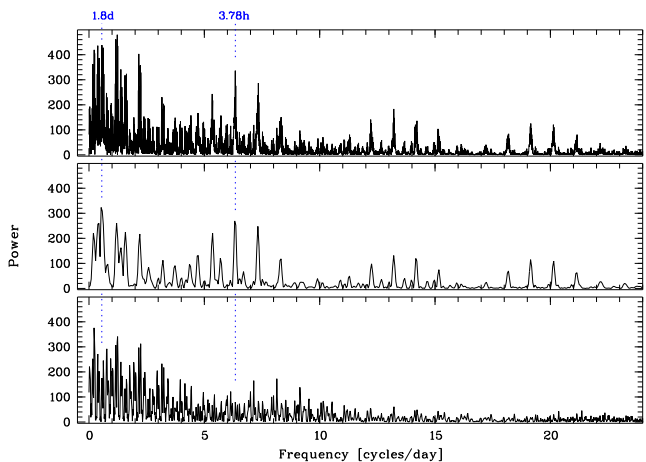}}}
\caption{\label{scargle_diff} The Scargle--diagram for the data minus the 
average light curve. Top diagram: all data, middle diagram: data until Feb 14,
lower diagram data from March and April. 
A new peak becomes visible indicating a period of 3.78\,h which is present in
the first part of the data but not in the later observations.}
\end{figure}

For the further analysis, we subtracted from each data point the corresponding
value of the average light curve. The resulting residual light curves
are plotted in Fig.\,\ref{diff_all}.
As expected they are dominated by the strong flickering. 
However, there seems to be an additional
brightness variation present on longer time--scales.
We checked the periodograms for any periodic signal and indeed 
found a peak at $f_{\rm sh} = 6.34$\,cycles/day, which corresponds to a 
period of 
$P_{\rm sh} = 3.78$\,h (see Fig.\,\ref{scargle_diff}). 
The interpretation of this period as the superhump period is supported
by the presence of various typical frequencies in the numerology of superhumps
as described e.g. by \cite{patt+05}. We find peaks at 0.56, 13.21, and 20.14
cycles/day, which correspond to  $\Omega = 1/P_{\rm orb} - 1/P_{\rm sh}$,
$2f_0 - \Omega$ and $3f_0 - \Omega$. The beat period 
$1/\Omega = 1.79$\,d is thus interpreted as the probable precession 
period of the accretion disc.

\begin{figure}
\centerline{\rotatebox{-90}{\resizebox{!}{8.3cm}{\includegraphics{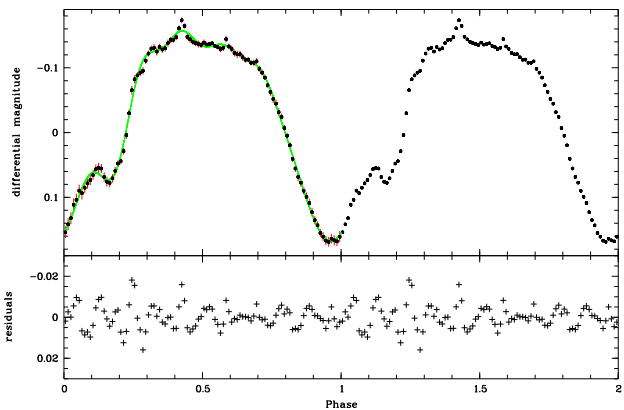}}}}
\caption{\label{ave_res} The average differential V--magnitude of RR\,Pic
is plotted against the phase using the period $P = 0.1450$\,d. Two cycles
are plotted for clarity.
The zero-phase was chosen to correspond to the eclipse--ephemeris as defined
by \citet{schm+05}. For the first cycle, error bars representing the sigma
of each point are over--plotted,
and the fit using the orbital period and its harmonics (see text for details) 
is given as a line.
Below, the residuals of fit and average light curve are plotted. }
\end{figure}

A detailed investigation on the robustness of this peak is not as successful
as for the orbital period. While the peak is present in all combinations of
data taken in February, it does not appear in the data taken in March or April.
This might in part be explained by the way the observations were performed.
The February observing run was dedicated to RR\,Pic, which was thus 
observed for at least two orbits every night. In March and April, the 
observations were done in service mode, and in general, only slightly 
more than one orbit was observed per night. 

\begin{figure}
\centerline{\rotatebox{-90}{\resizebox{!}{8.3cm}{\includegraphics{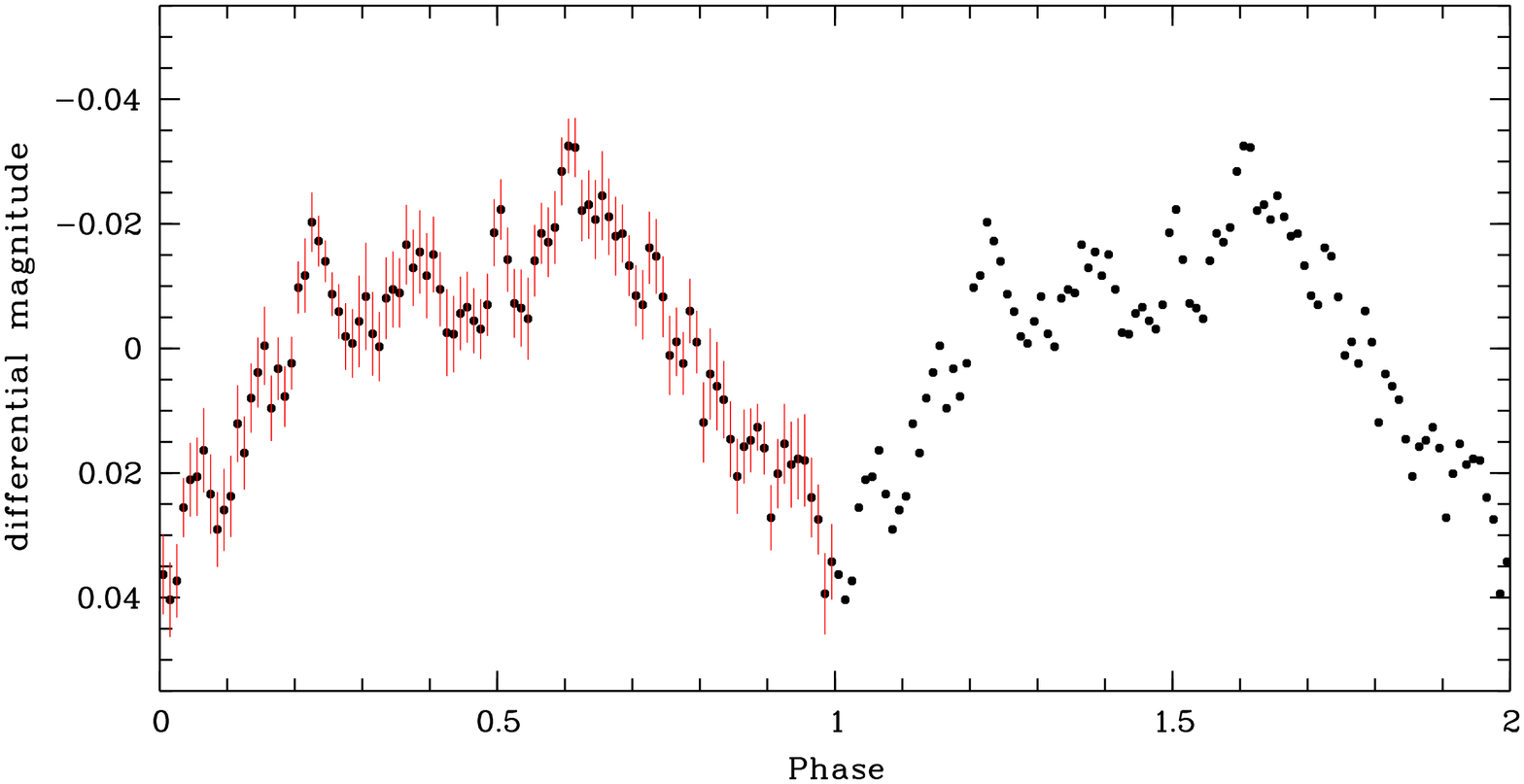}}}}
\centerline{\rotatebox{-90}{\resizebox{!}{8.3cm}{\includegraphics{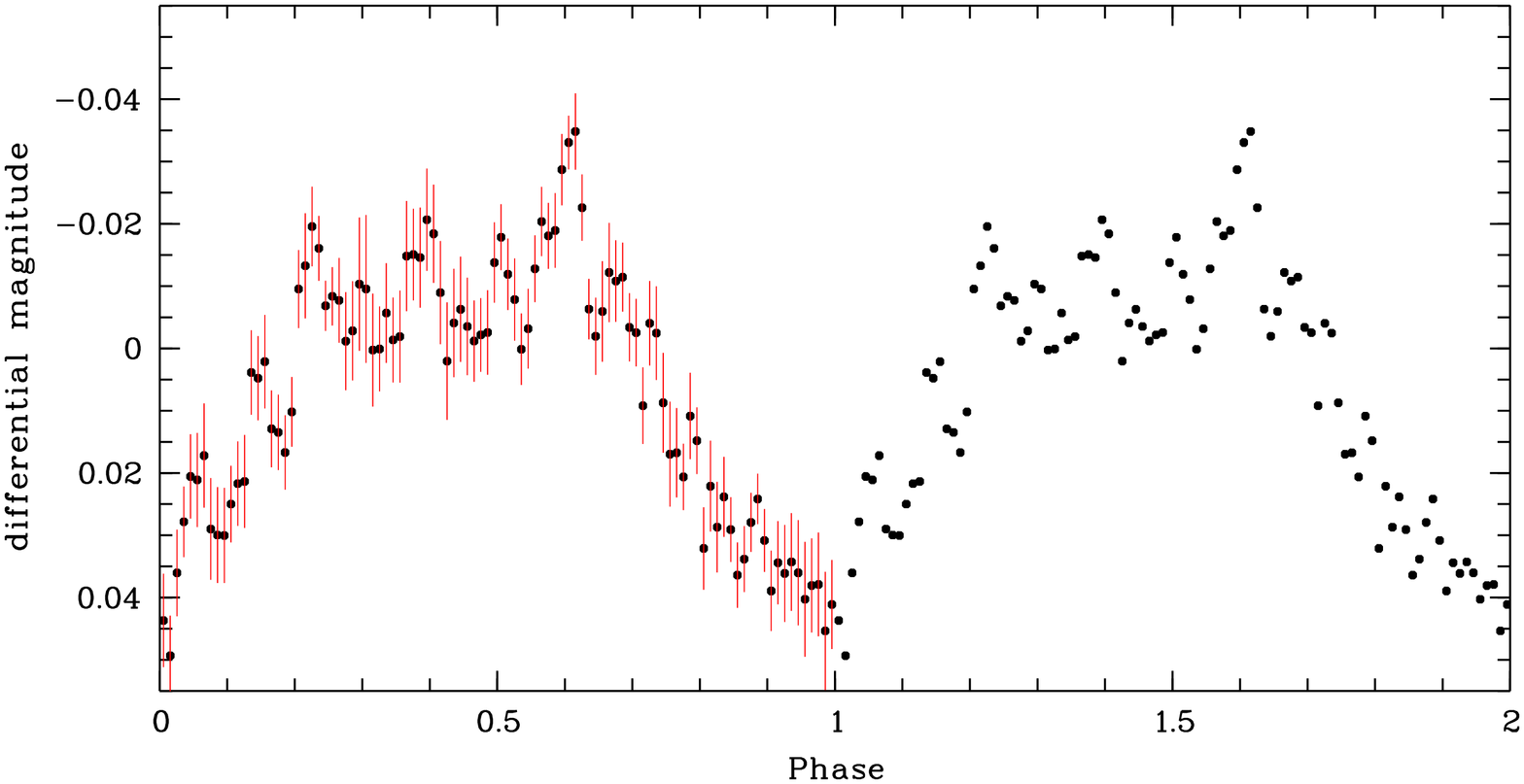}}}}
\centerline{\rotatebox{-90}{\resizebox{!}{8.3cm}{\includegraphics{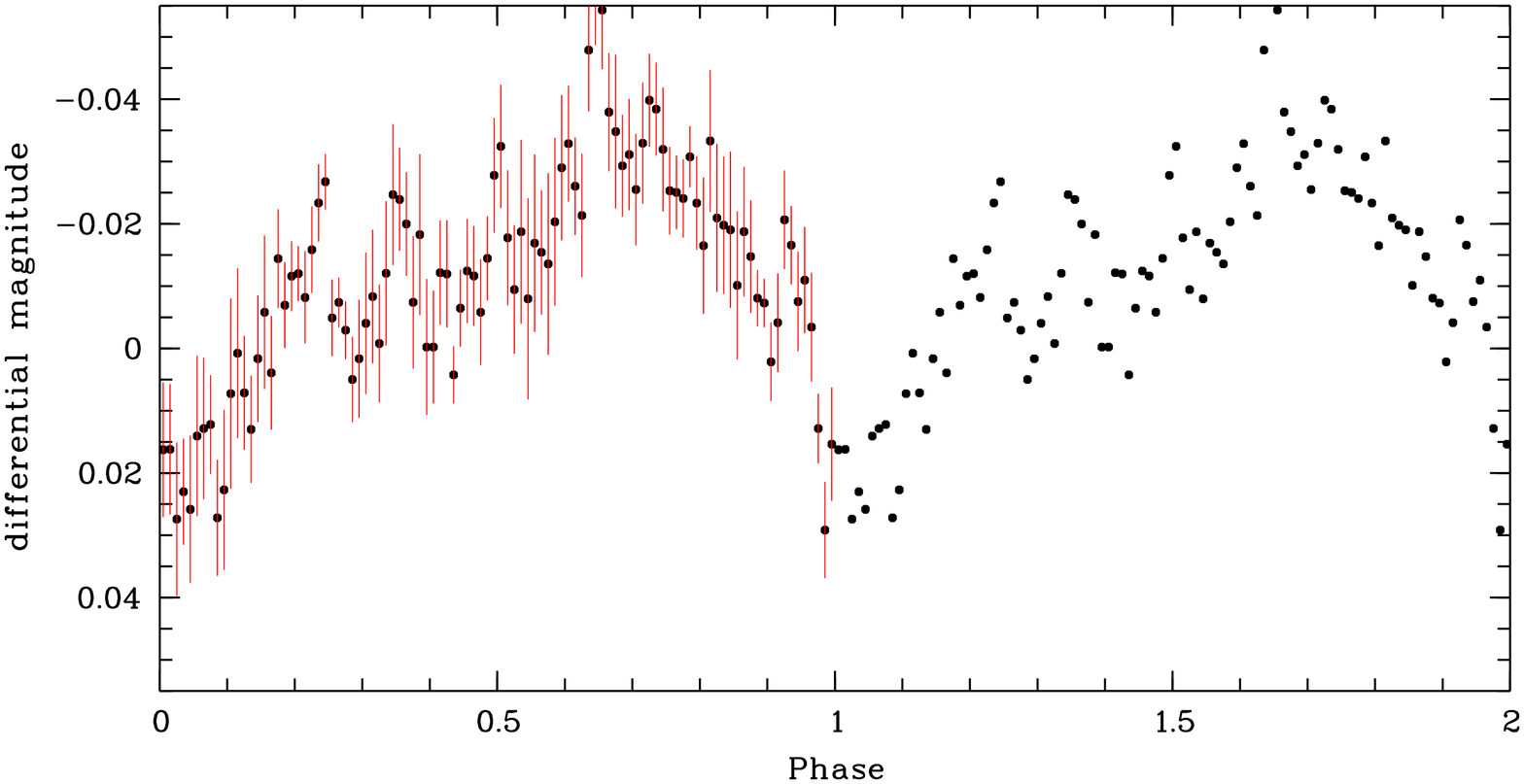}}}}
\caption{\label{ave_diff} The average differential V--magnitudes of RR\,Pic,
 of which the average orbital variation was subtracted, are plotted against 
the phase using the period $P_{\rm SH} = 0.1577$\,d. Top diagram: all data, 
middle diagram: data until 14th February, lower diagram: data from March and April. The superhump is clearly visible in all plots.}
\end{figure}

We averaged the residual light curve for the different data sets 
using the period of the superhump.  
In Fig. \ref{ave_diff}, these average residual
light curves are plotted. They clearly show
that the superhump is present during all our observing runs. 
The absence of a clear corresponding peak in the
Scargle--diagram is thus only due to the observing strategy and does not 
indicate the absence of the superhump in the latter data.

Fig.\,\ref{ave_diff} also gives the impression 
that the superhump is highly structured. Short--periodic variations are present
especially during the early phases (0--0.7), while the later part
of the lightcurve looks rather stable. The amplitude of the high--frequency
variations is around 0.01\,mag, thus it is similar in frequency and amplitude
to the variations seen in the residuals of Fig.\,\ref{ave_res}.

\subsection{Quasi--periodic oscillations (QPOs) and flickering}
\begin{figure}
\rotatebox{-90}{\resizebox{!}{8.3cm}{\includegraphics{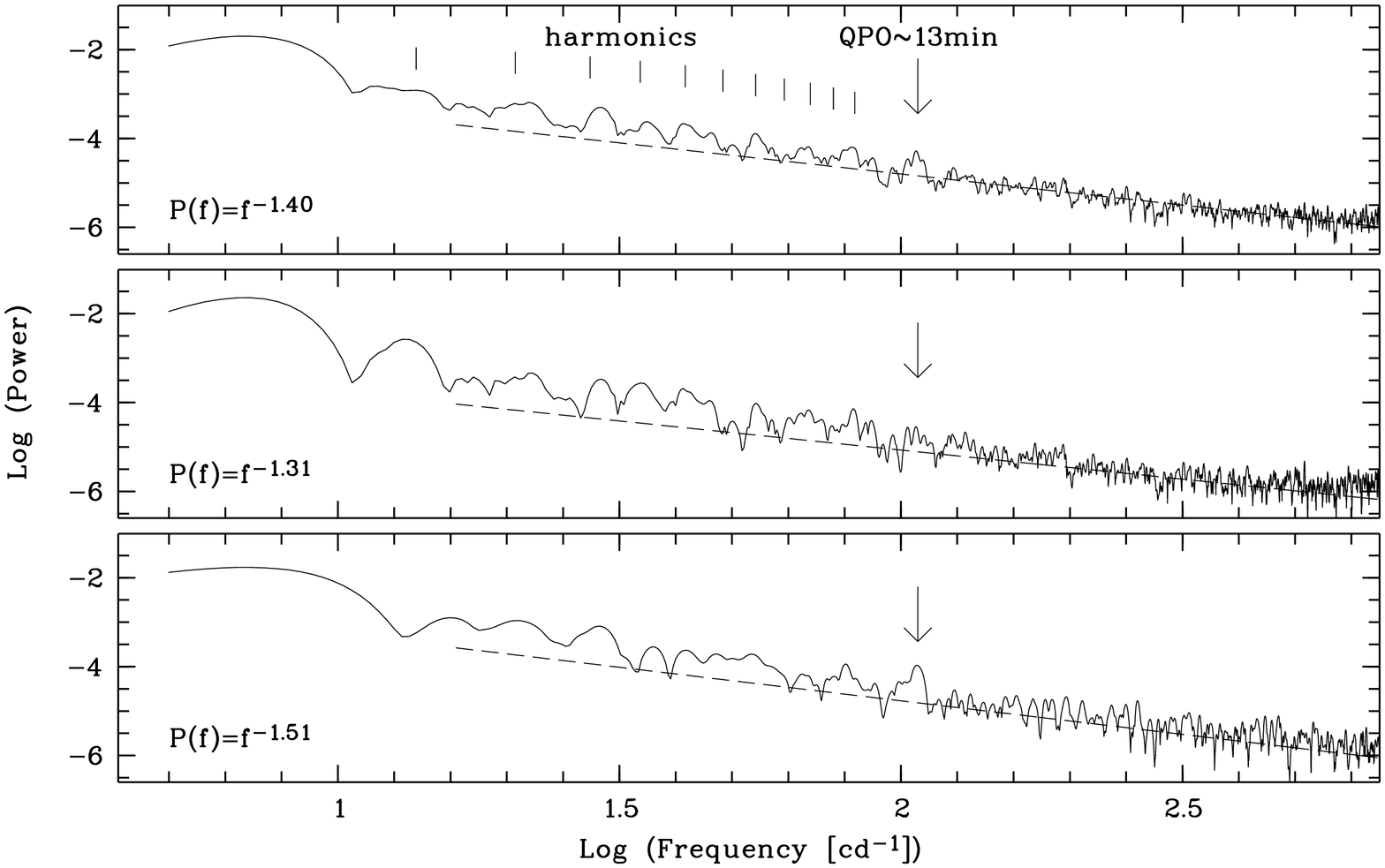}}}
\caption{\label{QPOs} Average power spectra in logarithmic scale of RR\,Pic's 
light curve. Top diagram: all data, middle diagram: data until 14th February, 
lower diagram: data from March and April. Indicated are the positions for the
harmonics of the orbital period up to 11th order and the additional bump at 2.04 which we attribute to a QPO feature.
The fits to the linear parts are overplotted, the corresponding equation 
given at the lower left part of each plot.}
\end{figure}

High--frequency variations present in Fig.\,\ref{lc_all} and 
Fig.\,\ref{diff_all} suggest the possible presence
of QPOs. 
The time--scale of these variations is
in the order of 15\, min, and they
seem to repeat themselves albeit with inconsistency in amplitude
and frequency. We therefore
searched the periodograms for signals that might indicate the existence 
of possible QPOs in RR\,Pic. Although, we
did not detect any coherent signal, all nights do show 
several peaks in the power spectrum in the range between 90 and 130\,1/d.
As Fourier analysis can omit signals, such as QPOs, that are unstable in
amplitude and frequency (see \cite{papa+06} for such examples), we applied
the following procedure in order to enhance a possible QPO signal.
%
The power spectra of all
nights were averaged and plotted in log--log scale (see Fig. \ref{QPOs}).
In this plot, the harmonics of the orbital period can be followed down to
the 11th harmonic at 82.8\,1/d, whose power exceeds the continuum by a 
factor of 1.9. Around $\log{F} = 2.04$\,1/d a broad peak becomes visible,
pointing to the counterpart of a QPO as suspected from the inspection
of our light curves. The individual average power spectra of February and
March--April 2005 (middle and lower plot of Fig. \ref{QPOs}, respectively)
show that the 13\,min QPO is always there but more prominent in the
March--April data. 
The power of this feature exceeds the continuum by a factor of 4. It 
is the counterpart of a 13\,min 
oscillation which agrees with the typical value for the oscillations present
in the light curves and even more visible in the residual light curves (see Fig. 
\ref{lc_all} and \ref{diff_all}). Note that the peak is present in the
February data as well as the March--April data but is more dominant 
in the latter.

The flickering in CVs, described through a shot noise--like process,
results in the so--called ''red noise'' seen in the power spectra
of CVs as the exponential decrease of power with frequency.
It thus follows linearity in a log--log scale diagram.
If all shots have the same duration, the periodogram of the resulting
light curve equals the periodogram of a single shot and thus just describes its
shape. 
The power law index $\gamma$, given by the slope of the linear decrease in
the log--log scale, has a value of 2 in the classical ''shot noise'', 
where the shots have an infinite rise--time and then decay. 
If, however, as expected from mechanisms generating the flickering,
the shots' durations are different and follow some kind of distribution,
then $\gamma$ gets smaller \citep{bruch92}.  
In this way, $\gamma$ can be used for the
characterisation of flickering activity but so far has been unsuccessful
to advance on understanding the physical origin of flickering
activity.
For more detailed information on the mechanisms and 
the resulting power law, see \citet{papa+06} and references therein.


Clearly, Fig.\,\ref{QPOs} shows that RR\,Pic is also characterised by
''red noise''.
We fitted the linear part for frequencies above 100 c/d by
a least--square linear fit and determined $\gamma = 1.40(2)$
for the average log--log power spectrum.
Given the noisiness of the power spectrum, the precise choice of the 
fitting interval becomes more difficult. Applying small changes of this
interval causes changes in the value of $\gamma$ up to 0.13 and therefore
0.1 should be considered as the real uncertainty of $\gamma$.
Within this error, the flickering pattern remains stable
within the span of our observations. 

We checked the photometric data taken in 2004 \citep{schm+05} for the presence
of the QPOs and the flickering pattern. Only two orbits go into this 
analysis and the average power spectrum is rather noisy. We find no 
indication for the presence of a QPO. The slope of the linear
decrease in the log--log power spectrum has been determined as $\gamma = 1.6$.

\section{Discussion}
\subsection{The orbital period and the radial velocity phase shift}

With the data spanning two years, we derive an orbital period
of $P=0.1450255(1)$\,d which agrees well with the formerly reported ones
within our uncertainty.
We can thus only give an upper limit to any possible change of the 
orbital period as $10^{-7}$\,d or about 0.1s.
 
The formerly reported phase shift of 0.1 phases that was observed in the 
radial velocities of data taken two years apart \citep{schm+05},  
can thus not be explained by a change of the orbital period.
To obtain a shift of 0.1 phases after such about 5000 cycles, 
the period must have changed by at least
$\rm 0.1 \times 0.14502545d / 5000 =  3\cdot 10^{-6}d$, which is in 
contradiction to
the new measurements. This shift is thus best explained by a varying
emission structure in the accretion disc of RR\,Pic which can influence the 
shape and phasing of the radial velocity curve. Such variations have been
observed by \cite{schm+06} using Doppler tomography techniques on several
emission lines in the spectra of RR\,Pic. Also, the structural change 
suggested by \citet{warn86} supports the idea that the accretion disc
of RR\,Pic is not stable but undergoes changes of various extent.

\subsection{Superhumps and QPOs}
The interpretation of the newly found variation with a period of 
$P_{\rm sh}=3.78$\,h as a superhump seems obvious due to the presence
of the typical frequencies as described in section 3.2. However, it has to 
be noted that the resulting value for $P_{\rm sh} - P_{\rm orb}= 18$\,min
is rather large for a nova--like star and implies a rather large mass ratio. 
Still, such large mass ratios are not unlikely for high--mass transfer CVs.
\citet{patt01} relates the mass ratio $q$ to the orbital and superhump
period via an empirical formula
$\epsilon = 0.216 \cdot q$ 
with
$\epsilon = \frac{P_{\rm SH} - P_{\rm orb}}{P_{\rm orb}}$.
In the case of RR\,Pic, we find $\epsilon = 0.0860$ and thus
derive a mass ratio of $q = 0.39$. This value would actually be above the
critical mass ratio for which superhumps are observed and is unlikely to be
correct. In fact,
\citet{patt+05} revised the formula especially for large mass ratios
where the original one would predict too large values of $q$. 
Using this new formula 
$\epsilon = 0.18 q + 0.29 q^2$, we derive a mass ratio $q=0.31$ for RR\,Pic,
which is more reasonable.
This value is slightly higher than the value 
found by \citep{ribe+06} who used a mass diagram calculated from their 
radial velocity measurements to derive $0.1<q<0.2$. 
However they do
acknowledge the fact that the radial velocities might be strongly
influenced by emission sources in the accretion disc. Since we know 
that these are not stable, the radial velocity curve might not actually
trace the velocity of the white dwarf. Increasing the velocity of the
white dwarf would yield a higher mass ratio $q$. On the other hand,
the formula used above describes an empirical average and does not necessarily
give the mass ratio for individual systems, so caution is advised also 
with this method. To unambiguously determine the
masses involved in the RR Pic system, observations of the secondary star 
and its radial velocities are needed.

Looking at Figure \ref{ave_diff}, there seems to be some short--term variation
pattern stable with the superhump period. Thus the question arises whether
the QPOs are connected to the superhump phenomenon. However, zooming into 
the lightcurve folded on the orbital period one notices
a similar short--term variation (not shown). 
So both, the orbital variation as well as the superhump have sub-structures.
Note that the frequency of the found QPOs is not a harmonic of the superhump's 
main frequency, it 
rather lies right in between the 17th and 18th harmonic.

\citet{kubi84} had found some similar oscillations, which however were not 
confirmed afterwards. 
He reported a most likely period of 15 min, although  
13 or 17\,min were also possible with his data.
Comparing this information with Fig.\ \ref{QPOs}, 
we see a small peak at 15\,min, nothing at 17\,min and the maximum is 
clearly at 13min. From this, we conclude that we actually confirm
Kubiak's findings although we would place 
the most likely period at 13\,min rather than at 15\,min.
The fact that these QPOs were not present in our data from 2004 is consistent
with the fact that Kubiak's variations were not confirmed in later observations.
RR\,Pic seems indeed to change its behaviour every now and then and not all
observational phenomena are present at all times.

\subsection{Is RR\,Pic eclipsing?}
RR\,Pic has been reported by various authors to show 
eclipse--like features (\citet{warn86},
\citet{haef+91}), while others have not noted this 
(\citet{vogt75}, \citet{kubi84}). Vogt confirmed that he never saw any evidence
for the presence of an eclipse in his data (private communication). In fact,
\citet{warn86} compared the lightcurves taken during the 60s and beginning 
of the 70s with those taken later in the 70s and 80s and suggested a change
in the structure of the accretion disc to explain the different appearance 
of the lightcurves. While the early lightcurves were dominated by two humps
of about 0.3\,mag brightness
and several small minima, the later lightcurves were rather flat (no variation
larger than 0.1\,mag) and showed 
the already mentioned eclipse--like feature.

If we put our observations in this context, it seems as if RR\,Pic is
back to or at least closer to its state in the early 70s. 
We observe a hump-like feature of about 0.3\,mag brightness, 
we see the two minima reported by Vogt, a broad one 
followed by a smaller one, but we find no evidence for an eclipse.

Comparing the overall shape of the average light curve with previous 
observations (found e.g. at 'The Center for Backyard Astrophysics`\footnote{http://cbastro.org/cataclysmics/atlas/rrpic.html})
suggests that if the eclipse was present, it should actually
correspond to the small minimum that we observe at phase 0.2 and not
to phase 0 which would be the original eclipse phase reported by Warner.

Another interesting point is the fact that the minimum of the superhump
falls on the phase of the eclipse. Since no ephemeris were known for the superhump, we
arbitrarily set the 0-phase of the superhump lightcurve to Warner's eclipse phase.
This means that at least on this eclipse observed by Warner, the superhump 
feature was also in minimum. Maybe, the eclipse feature is in fact 
a resonance phenomenon of the minima of the orbital lightcurve and the superhump
lightcurve.  Such resonance phenomena have been observed before.
E.g. in the dwarf nova OU\,Vir, a clearly eclipsing system, shows
superhumps during outburst, in which case the deepness of the eclipse
varies between 0.4 and 1\,mag and is modulated with the precession cycle, i.e.
the beat of orbital and superhump period. Assuming that RR\,Pic has a lower 
inclination and in general more shallow eclipse, it's detection or 
non-detection could well be modulated with the 1.79\,d precession period.
\citep{patt+05}. 
Warner stated that the eclipse was shallow and not present 
in all cycles, so this might support the idea of a resonance amplification.
If the superhump minimum coincides with the second minimum in the lightcurve, it might
enhance this one to be taken for an eclipse as in the 
'The Center for Backyard Astrophysics` data. 

On the other hand, we know from Doppler tomography (\citet{schm+03}, \citet{schm+05} and \citet{ribe+06}) that structural changes do take place in the
accretion disc of RR\,Pic. As such, also the visibility of an eclipse might
be influenced by these changes. 
It would be 
interesting to combine photometric variability observations with Doppler
tomography of the same night to actually compare the appearance of the
accretion disc in the Doppler map with the shape of the lightcurve, and i.e.
the presence of an eclipse.

\section{Conclusions}

We have presented optical lightcurves of RR\,Pic and shown that they are 
dominated by a strong orbital variation. The orbital period derived from this
data is 
consistent with the previous reported ones. I.e. a change of this period
can not be responsible for the previously observed shift in the phases of
radial velocity curves. Instead, this shift is rather due to structural 
changes that are known to occur in the accretion disc of RR\,Pic. 

In addition to the orbital variation, a superhump is found that was used 
to derive the mass ratio $q = 0.31$. This value does not agree with a
previously reported lower one, further observations are needed for
clarification. 

QPOs of 13\,min are present in all our data taken between February and 
April 2005. Older data from 2004, do not show this oscillation. While  
our analysis thus confirms the variations reported earlier, it also shows
that these QPOs are a transient phenomenon.
Their presence might be connected with the accretion disc's structure if they
occur due to an illumination of blobs in the inner accretion disc from a 
spinning white dwarf.

From our data we can not confirm the presence of an eclipse. Instead, we
note that at least in one historical case the eclipse occurs when the minima of orbital
light curve and superhump lightcurve fall together. This might indicate that
the observed eclipse is a resonance phenomenon between the two lightcurves
and its existence or not-existence is modulated with the precession period.
We would like to clarify that we do not rule out the presence of a shallow
eclipse but insist that either its visibility is enhanced by the resonance
or a favourable structure of the accretion disc is needed for 
an eclipse to be observed.

In general we conclude that RR\,Pic is a highly variable system. The 
previously reported changes that happen in the accretion disc are probably
responsible for the 
various features in the lightcurve that are not present 
at all times. To really understand what is going on in this system, parallel
time--resolved spectroscopy and photometry would be needed over several cycles.

\section*{Acknowledgements}
We thank Rebeccah Winnick for doing a wonderful job in organising the
service observations.

\label{lastpage}
\appendix
\begin{figure*}
\centerline{\resizebox{7cm}{!}{\includegraphics{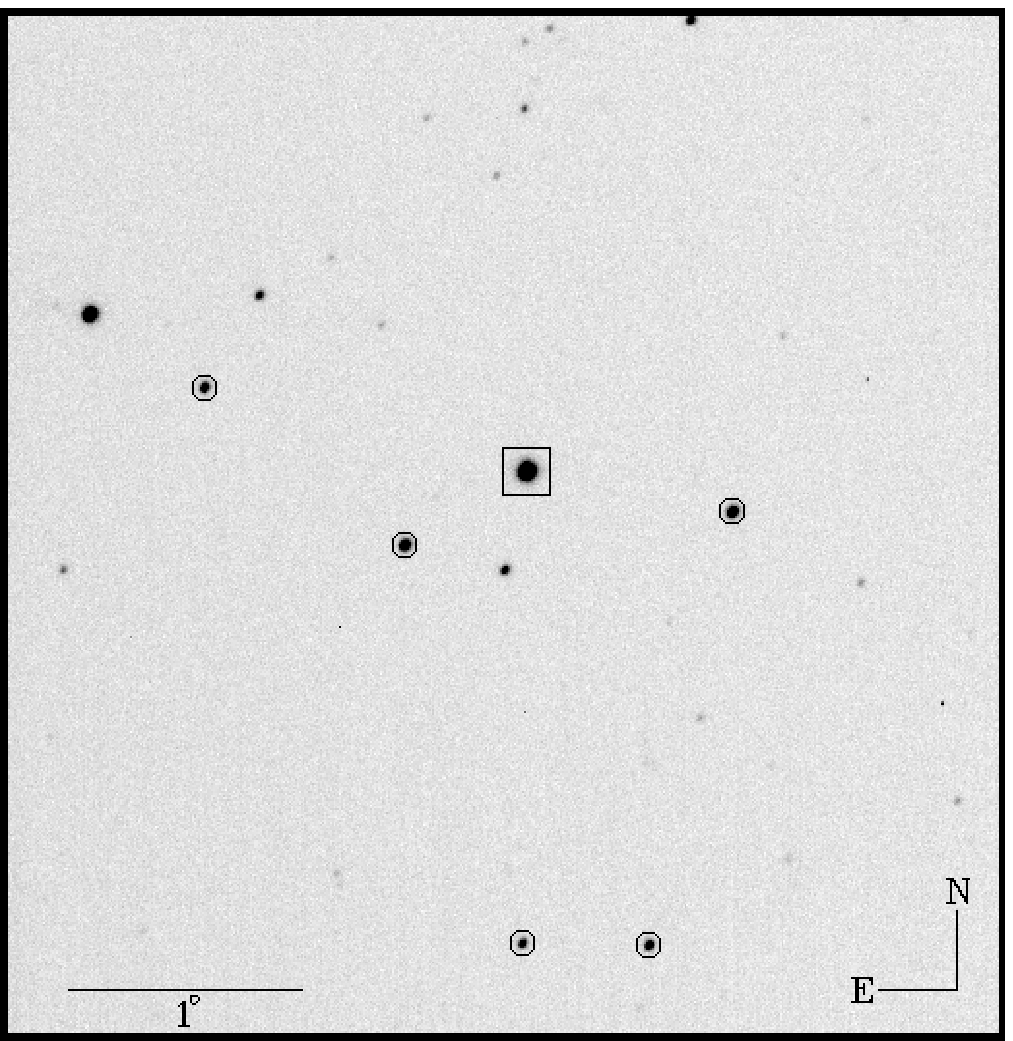}} ~ 
\resizebox{7cm}{!}{\includegraphics{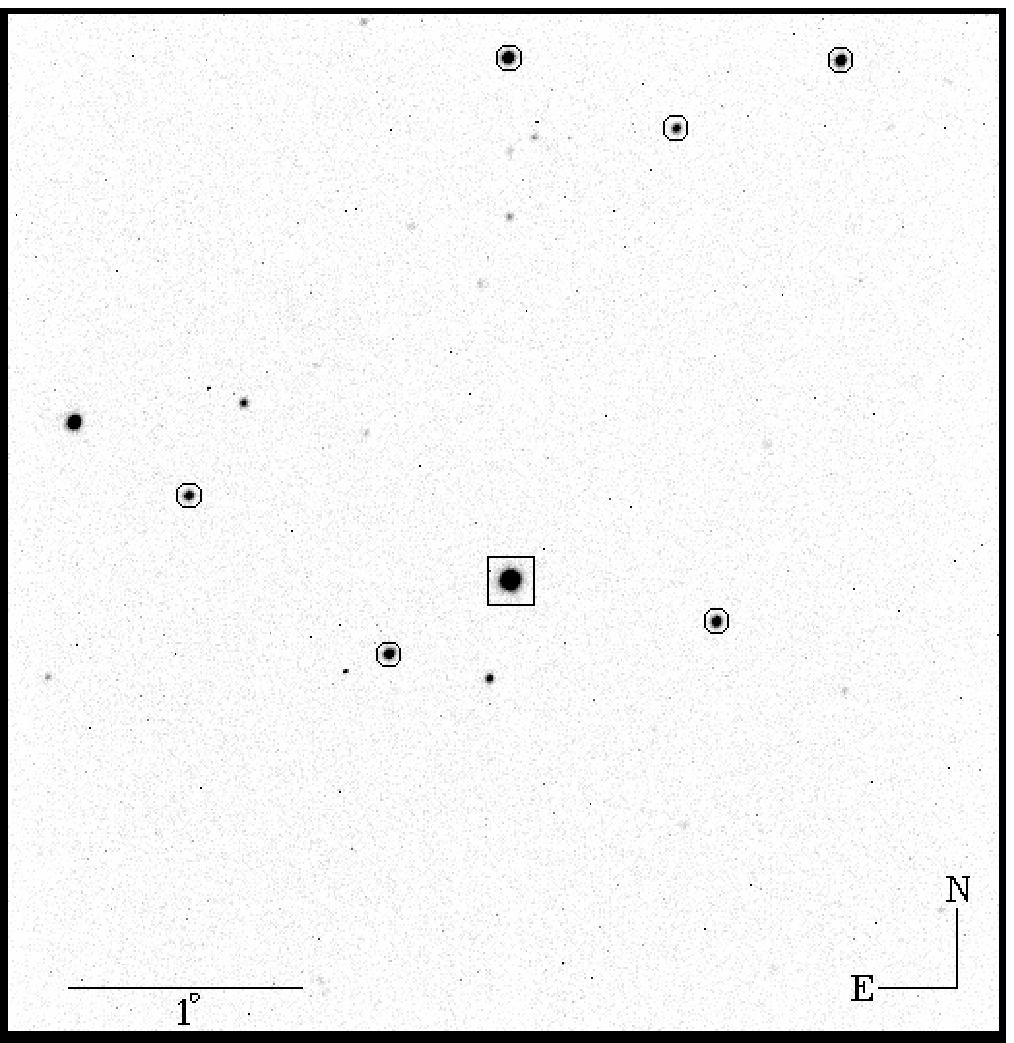}}}
\caption{\label{fc} The finding chart of RR\,Pic (indicated by the square in both images) with the two selected sets of
comparison stars (indicated by circles). The left image corresponds to ID1, the 
right one to ID2 (see table \ref{obstab} and text).}

\end{figure*}

\end{document}